\documentclass[aps, prb, twocolumn]{revtex4}
\usepackage{ams}
\usepackage{graphicx}

\usepackage{color}

\begin{document}

\title{Phase Diagram of Magnetization Reversal Processes in Nanorings}

\author{Wen Zhang}

\author {Stephan Haas}
\email{shaas@usc.edu}

\affiliation{Department of Physics and Astronomy, University of
Southern California, Los Angeles, CA 90089, USA}

\begin{abstract}
We investigate magnetization reversal processes of magnetic
nanorings. Using a recently developed efficient cartesian
coordinates fast multipole method, magnetization switching
phase diagrams are constructed for such structures. We find that the
dominant switching
mechanisms strongly depend on the relative spatial dimensions of the
rings, i.e. height, width and radius. Furthermore, they are found
to depend on a
characteristic exchange length scale, which is determined by the
competing short-range ferromagnetic and long-range dipolar
interactions in these systems. Detailed simulations allow us to
identify three novel switching mechanisms, which should be useful
for building high density storage systems.
\end{abstract}


\maketitle


\section {Introduction}

Mesoscopic magnetic elements have attracted tremendous attention
during the last two decades .\cite{De'Bell,Liu,Cowburnreview} In
particular, there has been an upsurge of interest in small magnetic rings
during the last few years.\cite{KlauiReview2003,KlauiReview2007} These
systems exhibit many novel physical phenomena with promise for
applications, such as high density data
storage\cite{cmuJAP2000,MRAM,360domain2}, magnetic
logics\cite{Cowburnlogic} and biomedical
sensoring\cite{sensor1,sensor2}. Furthermore, nanoring geometries
have proven to be a fertile platform for the investigation of
fundamental physical questions concerning domain
walls.\cite{KlauiReview2008} One of the most useful properties of
magnetic rings is that they can stabilize flux-closed vortex states.
These vortex states are free of stray fields and largely insensitive
to edge imperfections, thus resulting in highly controllable
magnetization reversal processes. Vortex states have been found to
be stable in rings with diameters as small as 10 nm. In contrast,
vortex states in magnetic discs exist only when their size is fairly
large, i.e. with diameters larger than 100 nm, due to the existence of
vortex core regions, which lead to high energy penalties. \cite{me}

Experimentally, nanomagnetic rings have been produced with a wide
range of spatial dimensions, extending from dozens of micrometers
down to nanometer scales\cite{ultrasmall1}. Various states have been
observed to be stable or metastable at remanence, i.e. in the
absence of an applied magnetic field, including the vortex state
(V), the onion state (O), i.e. two ferromagnetic domains separated
by domain walls at opposing ends of the ring structure, and the
twisted (or saddle) state (T)\cite{saddle}), i.e. a vortex state
interrupted by a $360^o$ domain wall. Several works have been
devoted to determining the phase diagram for nanoring structures
\cite{nanoring1,nanoring2,me,KlauiAPL2006}. Apart from the phase
diagram at remanance, magnetization reversal processes driven by
external magnetic fields are another important characteristic
property of magnetic elements. Several switching processes have
already been discovered \cite{meReview}. Experimentally, switching
phase diagrams have been constructed for micron-scale rings.
\cite{KlauiJMMM2005} However, the properties of nano-scale rings are
still under investigation, and a number of new features have
recently been discovered.\cite{RossPRB2003,RossJAP2005,ultrasmall2}

In this paper, we identify several switching processes, which
according to our knowledge have not yet been reported in the
literature, and construct switching phase diagrams for magnetic
nanorings. We use Monte Carlo simulations combined with a recently
developed efficient cartesian coordinates fast multipole
technique\cite{scale,mc1,mc2,mefmm} to analyze the magnetization
reversal processes for various nanoring dimensions. Neglecting the
crystalline energy, the total energy ($E$) of a magnetic
nanoparticle in a magnetic field consists of three terms: the
exchange interaction, the dipolar interaction, and the Zeeman
energy,
\begin{eqnarray}\label{sum}
E&=&-J\sum_{<i,j>}\vec{S}_i\cdot\vec{S}_j
+D\sum_{i,j}\frac{\vec{S}_i\cdot\vec{S}_j-3(\vec{S}_i\cdot
\hat{r}_{ij})(\vec{S}_j\cdot \hat{r}_{ij})}{r_{ij}^{3}} \nonumber
\\& & - \vec{H}\cdot \sum_i \vec{S}_i .
\end{eqnarray}
Here $J>0$ is the ferromagnetic exchange constant, which is assumed
to be non-zero only for nearest neighbors, $D$ is the dipolar
coupling parameter, and $\vec{r}_{ij}$ is the displacement vector
between lattice sites $i$ and $j$. Note that $\vec{S}$ is a dimensionless
unit spin vector with magnetic moments $\vec\mu=|\mu|\vec{S}$.

\begin{figure}[h]\begin{center}
\includegraphics[height=.4\textheight]{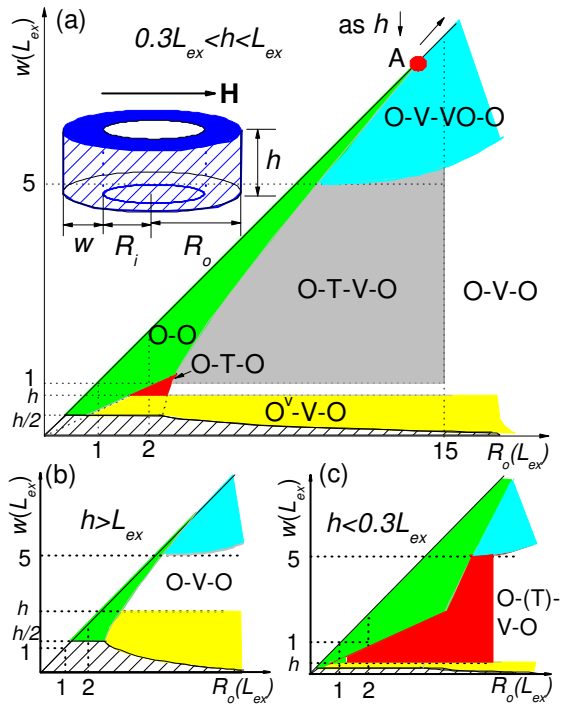}
\caption{\label{phase}(color online) Switching phase diagrams of
magnetic nanorings. As indicated in the inset, $h$ denotes the ring
height, $R_i$ and $R_o$ are the inner and outer ring radii, and
$w=R_o - R_i$ is the width. A magnetic field $\mathbf{H}$ is applied
in plane horizontally. The magnetic exchange length $L_{ex}$ is
defined in the text. Phase diagrams are shown for the cases (a)
$0.3L_{ex}<h<L_{ex}$, (b) $h>L_{ex}$, (c) $h<0.3L_{ex}$. The shaded
areas in the regime of small $w$'s represent rings whose remanence
states are out-of-plane ferromagnets, and hence not relevant for
switching. The color filled areas represent the different switching
processes discussed in the text.  green: onion - onion
(\textbf{O-O}); red: onion - twisted - onion (\textbf{O-T-O});
yellow: vertical onion - vortex - onion
(\textbf{O$^\textmd{v}$-V-O}); gray: onion - twisted - vortex -
onion (\textbf{O-T-V-O}); cyan: onion - vortex - vortex core - onion
(\textbf{O-V-VC-O}); white: onion - vortex - onion (\textbf{O-V-O}).
The point `A' indicates the onion - vortex transition for discs.
This critical point moves towards larger ($w,R_o$) values when $h$
decreases. }
\end{center}\end{figure}

Experimentally, the most studied magnetic nanomaterials are made of Co,
Fe, permalloy and supermalloy. For these materials, the ratio $D/Ja^3$
falls into the range between $10^{-3}$ and $10^{-5}$, where $a \simeq
0.3$ nm is the lattice constant. In the Monte Carlo simulations, single
spin updates are adopted to yield quasi time dependent behavior.\cite{mc3}
We choose $Ja^3/D=5000$ which is close to the parameters of Co. Since we
are trying to discover the generic property of nanomagnets, the absolute
values of parameters used in this paper are not as important as their
relative magnitudes. To state our conclusions in the most general terms, we
define an exchange length,
\begin{eqnarray}
L_{ex}=\alpha a\sqrt{Ja^3/D},
\end{eqnarray}
whose overall prefactor $\alpha$ may vary between 0.5 and 2,
depending on the material. Here we choose $\alpha = 1$, and assume a
typical value $a=0.3$ nm, such that $L_{ex}\sim20$ nm. This exchange
length is very important in determining the properties of
nanomagnetic systems. It is a measure of the spatial extent of
magnetic domains. For instance, the typical diameter of vortex cores
has been shown to be $1.3L_{ex}$.\cite{me} The temperature for the
Monte Carlo simulations is set to 50K, and at least $10^4$ Monte
Carlo steps are performed for each field point. To facilitate the
following discussion, we define $R_o$ as the outer ring radius,
$R_i$ as the inner ring radius, $w = R_o - R_i$ as the width and $h$
as the ring height (see the inset of Fig.\ref{phase}(a)).

\section {Switching Phase Diagrams}

We begin by summarizing our findings along with previously reported
results in the
switching phase diagrams shown in Fig.\ref{phase}.
Until now, there are three dominant magnetization
reversal processes which have been intensively studied and discussed in
the literature.

\subsection{Previously Reported Switching Processes}

\begin{figure}[h]\begin{center}
\includegraphics[height=.25\textheight]{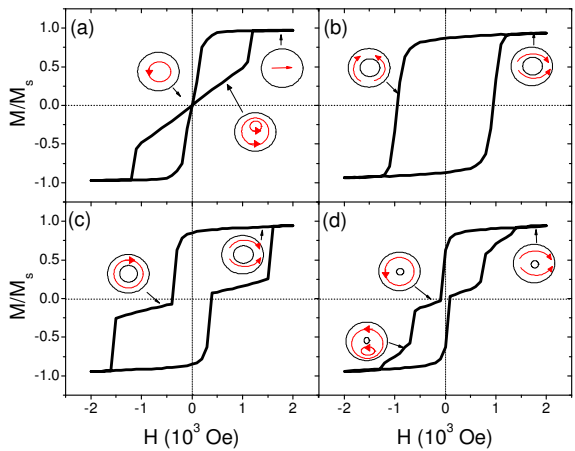}
\caption{\label{switching}(color online) Magnetic switching processes
in nanoring structures
and their associated hysteretic behavior. (a) hysteresis of magnetic
discs ($R_i \rightarrow 0$), (b) one-step \textbf{O-O}
switching of rings, (c) double switching \textbf{O-V-O} of rings, (d) triple
switching \textbf{O-V-VC-O} of rings. }
\end{center}\end{figure}

\begin{enumerate}

{\item  The first process is one-step (or single) switching
\textbf{O-O}, whose magnetization curve $M(H)$
is illustrated in Fig.\ref{switching}(b)). This is a
direct onion-to-onion switching process, whereby an onion state is
reversed to the opposite onion state directly via a coherent
rotation, as two magnetic domain walls move in the same rotational
direction. It is worth noting that
when the ring radius is
sufficiently small ($R_o<L_{ex}$),
the onion state is replaced by a
single domain ferromagnetic state.}

{\item  The second process is two-step or double switching
\textbf{O-V-O}, shown in Fig. \ref{switching}(c). Here, two opposite
onion states are separated by a vortex state. This is the dominant
switching process for most reported magnetic rings, stabilized by any
spurious spatial asymmetry. At the first transition point, either
one wall is pinned stronger than the other or both walls move
towards each other due to asymmetry. In both cases the two domain
walls approach each other with increasing applied magnetic field.
When these two domain walls meet, they annihilate each other,
provided the ring is sufficiently wide ($w>5L_{ex}$). If not,
intermediate metastable twisted states have recently been found to
exist for small narrow rings \cite{RossPRB2003}. In most cases, the
remanence state is an onion state.
However, when the ring is sufficiently thick and
wide\cite{francePRL2001,triple}, the remanance state
is a vortex state.}

{\item  The third process is triple switching \textbf{O-V-VC-O},
shown in Fig.\ref{switching}(d). \cite{KlauiAPL2004,triple,triple2}
The switching mechanism is essentially the same as for
\textbf{O-V-O}, except that the vortex state does not deform into an
onion state abruptly, but first nucleates a vortex core in one arm
of the ring. The core then moves slowly to the outer rim. During
this part of the
process the state is called a ``vortex core" state (\textbf{VC}). To some
extent, this is similar to the switching process of magnetic discs,
where the core in the vortex state is shifted by the magnetic field
and ultimately exits at the boundary (see Fig. \ref{switching}(a)).}
\end{enumerate}

These magnetization reversal processes are found in relatively
large rings ($R_0 \gg L_{ex}$) with spatial dimensions
in the micrometer range. However, it
is reasonable to expect that in the sub-micron and truly nano ranges, new
phenomena exist and wait to be explored. Therefore, we have performed
simulations for a wide range of parameters and identified the
following three additional magnetization reversal processes.

\subsection{Vertical Onion - Vortex - Onion Switching:
\textbf{O$^\textmd{v}$-V-O}}

\begin{figure}[h]\begin{center}
\includegraphics[height=.25\textheight]{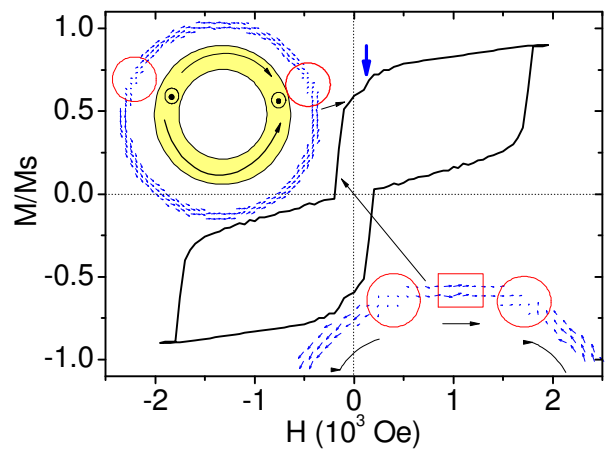}
\caption{\label{s1}(color online) Typical hysteresis curve for
out-of-plane \textbf{O$^\textmd{v}$-V-O} switching ($R_o=70$ nm, $w=7$ nm,
$ h=10$ nm). The insets are the snapshot during the \textbf{O$^\textmd{v}$-V}
transition. The
areas with strong out-of-plane components are highlighted by open red
circles. The big blue arrow points to a small step indicating the
appearance of out-of-plane domains.}
\end{center}\end{figure}

On first sight, this process appears to fall in the same category as
\textbf{O-V-O}, since its hysteresis, shown in Fig. \ref{s1}, shares
characteristics similar to double switching
(Fig.\ref{switching}(c)). The switching mechanism behind it,
however, is quite different. The onion state here does not consist
of two transverse head-to-head domain walls, but rather of two
vertical head-to-head domain walls (see the areas highlighted by
open red circles in Fig. \ref{s1}), where the spins point
out-of-plane rather than in-plane. Hence, we call the state
involving vertical head-to-head domain walls a ``vertical" onion
state (\textbf{O$^\textmd{v}$}). The big blue arrow in Fig. \ref{s1}
points to a small step in the magnetization curve, indicating the
appearance of out-of-plane domains. The two domains may point
parallel or antiparallel. However, the parallel configuration is
preferred, since the switching of \textbf{O-V} is more controllable
and the transition is sharper compared with the antiparallel case.
As we can see from Fig. \ref{s1}, when the two domains point in the
same direction, the \textbf{O-V} transition is simply achieved by
rotating spins out-of-plane in the intermediate region between the
two domain walls (highlighted by the red rectangle), which
subsequently align  with the out-of-plane domain wall spins. In the
antiparallel case, the domain wall which moves faster
will first rotate to form a normal in-plane head-to-head domain
wall. Then the other domain wall rotates to in-plane alignment, such
that the two domain walls can easily annihilate. This second process
results in a slightly larger and broader transition field, which
however can be easily avoided by applying a small out-of-plane
magnetic field.

\begin{figure}[h]\begin{center}
\includegraphics[height=.25\textheight]{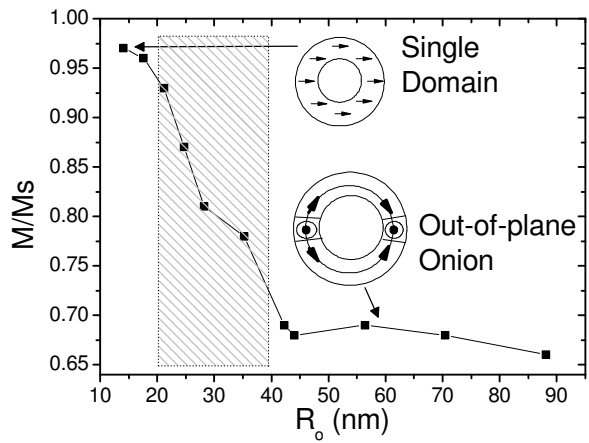}
\caption{\label{s11}(color online) Single domain to onion state
transition with incrreasing ring diameter. Here, the
height is fixed at $23$ nm, and the width is fixed at $15$ nm.
The shaded area represents the transition region. }
\end{center}\end{figure}

The \textbf{O$^\textmd{v}$-V-O} process has not been reported before,
as most of the initial
attention in the field had been focused on thin film rings ($h \ll w$).
As seen in the phase diagrams of Fig. \ref{phase},
this switching process indeed belongs to rings
with $w < h$ (see the yellow regions). Shape
anisotropies force the spins in the domain walls to point
out-of-plane. This process is crucial when one wants to use vortex
states in high-density memory applications. In the thin film case
($h<L_{ex}$), vortex states are not easy to form when the ring
radius is small (say $R_o<100nm$). Instead, twisted states will
stand in the way, preventing the formation of vortex states, which
will be discussed in detail next. The only limit of this
\textbf{O$^\textmd{v}$-V-O} process is the single domain limit for
the ring. As long as onion states exist, this scheme applies. However,
when
the ring is extremely small, the onion state will be replaced by a
single domain state. Roughly speaking, this transition happens at
$R_o\sim L_{ex}$. We use the remanence magnetization as an order
parameter to locate this transition, as shown in Fig. \ref{s11}. The
onion state is characterized by a relatively small magnetization
(approximately 0.68 for the chosen $w$ and $h$), since most of the
spins are aligned along the ring boundary, whereas the single domain
state is characterized by a magnetization close to saturation, since
in this case nearly all the spins align in the same direction.
Obviously, the transition region (shaded area in Fig. \ref{s11}) is
$R_o\approx 20-40$ nm $ \approx 1-2L_{ex}$. For rings with $R_o<L_{ex}$, the
only magnetization reversal process will be one-step switching via
coherent single domain rotation. Therefore, the $R_o$ of rings has to
be at least $L_{ex}$ if the vortex state is desired.

\begin{figure}[h]\begin{center}
\includegraphics[height=.25\textheight]{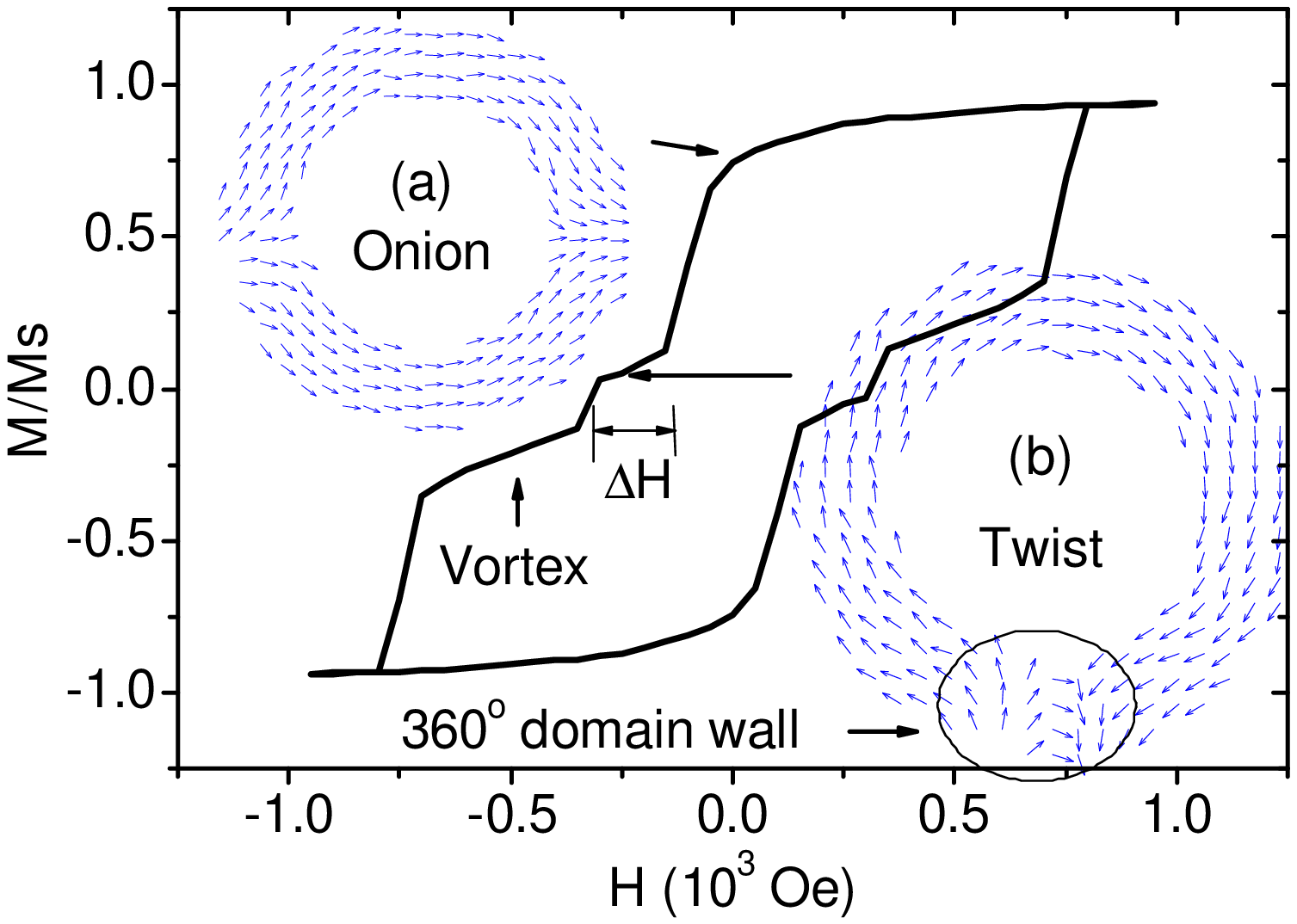}
\caption{\label{s2}(color online) Typical hysteresis curve for
quadruple \textbf{O-T-V-O} switching ($R_o=150$ nm,$ w=60$ nm, $h=15$ nm).
The insets (a) and (b) are snapshots during the transition. (a)
is the onion state at remanence, while (b) is a typical twisted
state, characterized by a $360^o$ domain wall.}
\end{center}\end{figure}

\subsection{Quadruple Switching: \textbf{O-T-V-O}}
This switching process consists of four steps. In contrast to
regular \textbf{O-V-O} triple switching, an additional step,
characterized by the appearence of a twisted state (see Fig.
\ref{s2} inset (b)), is involved. As mentioned above, twisted states
are only found in rings with relatively small lateral dimensions
\cite{RossPRB2003} ($R_o<500nm$). In order to understand the
conditions for the twisted state to occur, first recall the process
of onion-to-vortex (\textbf{O-V}) transition during the regular
triple switching: two domain walls move towards each other and
annihilate by introducing a vortex core passing through the rim,
i.e. entering from the outer edge and exiting through the inner
edge. As the diameter of the vortex core is typically equal to
$L_{ex}$, the width of the ring needs to be at least of the order of
several exchange lengths
in order to accommodate such a process. However, when the ring width
is sufficiently small, the annihilation process is hindered. In that
case, the two domain walls are stuck together, forming a $360^o$
domain wall (see Fig. \ref{s2} inset (b)). The configuration
involving this $360^o$ domain wall is called ``twisted
state".\cite{RossPRB2003} It turns out that this state can be
stabilized within a fairly large field range
$\Delta H$, e.g. for the specific
ring shown in Fig. \ref{s2} it is stable between 100 Oe and 300 Oe.
This result is consistent with previous literature\cite{RossPRB2003,
ultrasmall2}. Moveover, the results in Ref. \onlinecite{RossPRB2003}
show precursor behavior of this type of switching, since the vortex
state will not give positive magnetization when the magnetic field
is negative.


From our simulations we find that the field range $\Delta H$ of the
twisted state is sensitive to the geometric parameters ($R_o$, $w$,
$h$) in the following ways.

\begin{enumerate}
{\item $h<L_{ex}$: We find that the twisted state is not stable when
the ring height is larger than $L_{ex}$. It has been reported that
magnetic domain structures vary significantly along the vertical
direction when the height is large.\cite{vertical} Our
simulations show that this is indeed the case when $h>L_{ex}$, as the
exchange length $L_{ex}$ is a measure of typical domain size. In this
case, the vertical stacking of slightly off-set domains eases
the introduction of vortices in the annihilation process,
destabilizing the $360^o$ domain wall. Furthermore, $\Delta H$ is
most sensitive to ring height. The flatter the system, the larger
$\Delta H$, i.e. $\Delta H$ can be doubled with decreased height. On
the other hand, vortices cannot be formed if $h$ is too small
($<0.3L_{ex}$), which leads to the last switching process we will
discuss in the next subsection.}

{\item  $L_{ex}<w<5 L_{ex}$: Typically, the field range $\Delta H$
varies with ring width (about $20\%$), with a peak around $w=3L_{ex}$.
The reason for this is that there is a qualitative difference
between thin and wide rings when the $360^o$ domain wall is
destroyed. For rings with $w<L_{ex}$, the $360^o$ domain wall is not
stable, with the exception of ultra-small rings ($R_o<2L_{ex}$). As
it is the case for out-of-plane \textbf{O-V-O} switching, the spins
in the domain area first point out-of-plane, and then turn in-plane
along the field
with increasing applied field. Slightly wider rings can support the
$360^o$ domain wall in a small, but finite, field range and then
evolve with the same process. For wider rings with $w>5L_{ex}$, a
vortex can be introduced through the outer edge of the ring and
traverse through the rim, as already described for the
\textbf{O-V-O} triple switching case. Slightly thinner rings still
feature vortices. However, in this case the vortices cannot remain
in the area of the ring where they entered when the reversed
transverse field is increased.  Instead, they move along the rim of
the ring and finally exit. Hence, in this case the associated
magnetization step in the hysteresis curve
has a fairly large slope, indicating that the twisted region
is moving with increasing applied field. The competition of these
two mechanism gives rise to a peak of the field range where the
twisted state is stable, which occurs at $w \approx 3L_{ex}$.}

{\item $L_{ex}<R_o \alt 15L_{ex}$: In order to observe the twisted state,
the entire ring has to be sufficiently small, but still larger than
the single domain limit.
The reason why  larger rings destabilize the twisted state more easily is
that in these structures vortices can enter from the
outer rim. In smaller
rings, vortices are not as easily accomodated in the domain wall region,
because of differences between the arclengths of the outer
and inner domain wall boundaries, due to the increased curvature.
Another reason is that
for large rings, there is a higher chance that defects serve as a
catalyst of local vortex formation. $\Delta H$ is almost independent
of $R_o$ for fixed aspect ratios ($\kappa=R_i/R_o$). On the other
hand, the width of the magnetization step in $M(H)$
where the vortex state is stable
decreases quickly when $R_o$ is decreased, because the curvature
eases the process of introducing a vortex during the \textbf{V-O} process.
When $\kappa$ is large and the ring is ultra-small, the extreme case
where no vortex is involved emerges as discussed in detail below. }

\end{enumerate}

The above conditions are fulfilled in the gray regions
of the phase diagrams in Fig. \ref{phase}.

\begin{figure}[h]\begin{center}
\includegraphics[height=.3\textheight]{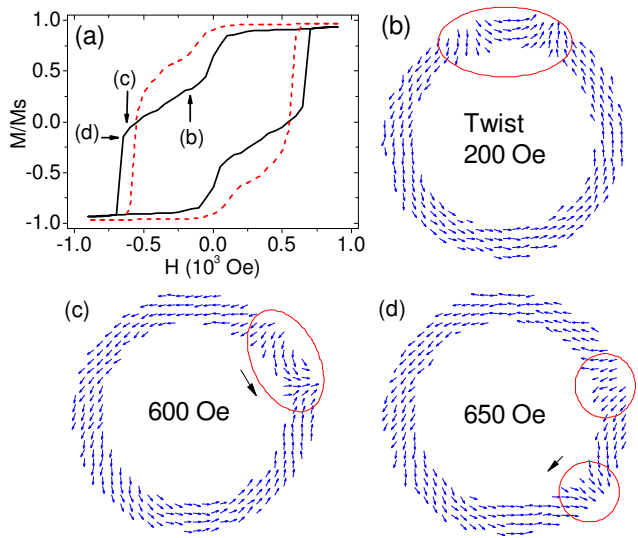}
\caption{\label{s3}(color online) Twisted triple switching
(\textbf{O-T-O}): (a) Hysteresis curves for ring 1 ($R_o=70$ nm,
$w=18$ nm, $h=5$ nm, black continuous line) and ring 2 ($R_o=42$ nm,
$w=16$ nm,$h=14$ nm, red dashed line); (b) spin configuration for ring 1
at $H=200$ Oe; (c)  spin configuration for ring 1 at $H=600$ Oe;
(d) spin configuration for ring 1 at $H=650$ Oe. }
\end{center}\end{figure}

\subsection{Twisted triple switching (\textbf{O-T-O})}
This process is an extreme limit of the \textbf{O-T-V-O} switching
process. Rather than forming a vortex state to lower the energy, the
system chooses to move the entire domain wall towards either end of
the ring along the magnetic field (see Fig. \ref{s3}(c)), when the
applied magnetic field is increased. Ultimately, the $360^o$ domain
wall breaks up into two $180^o$ domain walls, and one of them
travels back to the other side (see Fig. \ref{s3}(d)), forming the
reversed onion state. For relatively large rings, one can observe a
small step in the magnetization curve, indicating the occurence of
the twisted state (see Fig. \ref{s3}(b)), while this step is not
noticeable for very small rings.
After the twisted state is formed, the
magnetization decreases approximately linearly with increasing
applied magnetic field, indicating the movement of the domain wall
(see Fig. \ref{s3}(c)). Finally, a large jump is observed,
indicating the breakup (see Fig. \ref{s3}(d)) of the $360^o$ domain
wall and the formation of the reversed onion state. This process can
only be observed under two extreme conditions:

\begin{enumerate}

{\item Very small narrow rings (typically $R_o<3L_{ex}$ and
$\kappa>0.5$), when $0.3L_{ex}<h<L_{ex}$ (see the small red region
in Fig. \ref{phase}(a)). The red dashed line in Fig.\ref{s3}(a) is
the typical hysteresis in this category. Here it is the curvature of
the ring that prohibits the annihilation process. Under such
conditions, the domain region typically covers a quarter of the
ring.}

{\item  Ultra-thin small narrow rings (typically $h<0.3L_{ex}$ and
$\kappa>0.5$). The black solid line in Fig.\ref{s3}(a) is the
typical hysteresis. Fig.\ref{s3}(b)(c)(d) are snapshots of the
configuration evolution with increasing applied magnetic field.
In this case, it is the strong easy-plane
shape anisotropy that hinders the annihilation process. For these
ultra-thin systems, this switching process can be observed for rings
as large as $R_o\sim100nm$ (see the fairly large red region in
Fig.\ref{phase}(c)). Theoretically they may exist on even larger
rings. In reality, however, local defects will favor the formation
of a local vortices, destroying the $360^o$ domain wall.}

\end{enumerate}

Note that the twisted state is stable at remanence once it is
formed. Designs utilizing this state for magnetic memory have been
proposed\cite{360domain2}. Our results show that such designs can
use very small rings, for which there appears to be great potential
to build high density storage systems.

\section{Conclusions}

In conclusion, we have computed the phase diagrams of magnetic nanorings,
and identified three new types of magnetization
reversal processes, such that there are now six known types of
magnetization reversal processes in all reported for magnetic
nanorings. The \textbf{O$^\textmd{v}$-V-O} switching
features a new out-of-plane onion state and easy switching process.
Rings in this category have $h>w$, and $R_o$ can be as small as
$L_{ex}$. The quadruple switching (\textbf{O-T-V-O}) and twisted
triple switching (\textbf{O-T-O}) feature twisted states. Rings in
these two categories are fairly small (typically with outer radius
less than 200nm). Their width has to be less than about $4L_{ex}$ to
stabilize the $360^o$ domain wall. The field range where the twisted
state is stable is most sensitive to the ring height. Small heights
give rise to large field ranges, and there will be no twisted state
when $h>L_{ex}$. We have combined our findings with results from
previous literature into complete switching phase diagrams, shown in
Fig.\ref{phase}. These should be helpful in designing magnetic
nanorings with specific properties, and show the potential of
building high density storage systems based on ring structures.

\section{Acknowledgements}

We would like to thank N.T.
Jacobson and Y. Tao for useful discussions. Computing facilities
were generously provided by the University of Southern California
high-performance supercomputing center. We also acknowledge
financial support by the Department of Energy under grant
DE-FG02-05ER46240.\\

\bibliography{scale}

\begin{thebibliography}{33}
\expandafter\ifx\csname natexlab\endcsname\relax\def\natexlab#1{#1}\fi
\expandafter\ifx\csname bibnamefont\endcsname\relax
  \def\bibnamefont#1{#1}\fi
\expandafter\ifx\csname bibfnamefont\endcsname\relax
  \def\bibfnamefont#1{#1}\fi
\expandafter\ifx\csname citenamefont\endcsname\relax
  \def\citenamefont#1{#1}\fi
\expandafter\ifx\csname url\endcsname\relax
  \def\url#1{\texttt{#1}}\fi
\expandafter\ifx\csname urlprefix\endcsname\relax\def\urlprefix{URL }\fi
\providecommand{\bibinfo}[2]{#2}
\providecommand{\eprint}[2][]{\url{#2}}

\bibitem[{\citenamefont{De'Bell et~al.}(2000)\citenamefont{De'Bell, Maclsaac,
  and Whitehead}}]{De'Bell}
\bibinfo{author}{\bibfnamefont{K.}~\bibnamefont{De'Bell}},
  \bibinfo{author}{\bibfnamefont{A.~B.} \bibnamefont{Maclsaac}},
  \bibnamefont{and} \bibinfo{author}{\bibfnamefont{J.~P.}
  \bibnamefont{Whitehead}}, \bibinfo{journal}{Rev. Mod. Phys.}
  \textbf{\bibinfo{volume}{72(1)}}, \bibinfo{pages}{225}
  (\bibinfo{year}{2000}).

\bibitem[{\citenamefont{Martin et~al.}(2003)\citenamefont{Martin, Nogues,
  K.~liu, and Schuller}}]{Liu}
\bibinfo{author}{\bibfnamefont{J.~I.} \bibnamefont{Martin}},
  \bibinfo{author}{\bibfnamefont{J.}~\bibnamefont{Nogues}},
  \bibinfo{author}{\bibfnamefont{J.~L.~V.} \bibnamefont{K.~liu}},
  \bibnamefont{and} \bibinfo{author}{\bibfnamefont{I.~K.}
  \bibnamefont{Schuller}}, \bibinfo{journal}{J. Magn. Magn. Mater.}
  \textbf{\bibinfo{volume}{256}}, \bibinfo{pages}{449} (\bibinfo{year}{2003}).

\bibitem[{\citenamefont{Cowburn}(2001)}]{Cowburnreview}
\bibinfo{author}{\bibfnamefont{R.~P.} \bibnamefont{Cowburn}},
  \bibinfo{journal}{J. Phys. D: Appl. Phys.} \textbf{\bibinfo{volume}{33}},
  \bibinfo{pages}{R1} (\bibinfo{year}{2001}).

\bibitem[{\citenamefont{Kl$\ddot{a}$ui
  et~al.}(2003)\citenamefont{Kl$\ddot{a}$ui, Vaz, Lopez-Diaz, and
  Bland}}]{KlauiReview2003}
\bibinfo{author}{\bibfnamefont{M.}~\bibnamefont{Kl$\ddot{a}$ui}},
  \bibinfo{author}{\bibfnamefont{C.~A.~F.} \bibnamefont{Vaz}},
  \bibinfo{author}{\bibfnamefont{L.}~\bibnamefont{Lopez-Diaz}},
  \bibnamefont{and} \bibinfo{author}{\bibfnamefont{J.~A.~C.}
  \bibnamefont{Bland}}, \bibinfo{journal}{J. Phys.: Condens. Matter}
  \textbf{\bibinfo{volume}{15}}, \bibinfo{pages}{R985} (\bibinfo{year}{2003}).

\bibitem[{\citenamefont{Vaz et~al.}(2007)\citenamefont{Vaz, Hayward, Llandro,
  Schackert, Morecroft, Bland, Kl$\ddot{a}ui$, and et~al}}]{KlauiReview2007}
\bibinfo{author}{\bibfnamefont{C.~A.~F.} \bibnamefont{Vaz}},
  \bibinfo{author}{\bibfnamefont{T.~J.} \bibnamefont{Hayward}},
  \bibinfo{author}{\bibfnamefont{J.}~\bibnamefont{Llandro}},
  \bibinfo{author}{\bibfnamefont{F.}~\bibnamefont{Schackert}},
  \bibinfo{author}{\bibfnamefont{D.}~\bibnamefont{Morecroft}},
  \bibinfo{author}{\bibfnamefont{J.~A.~C.} \bibnamefont{Bland}},
  \bibinfo{author}{\bibfnamefont{M.}~\bibnamefont{Kl$\ddot{a}ui$}},
  \bibnamefont{and} \bibinfo{author}{\bibnamefont{et~al}}, \bibinfo{journal}{J.
  Phys.: Condens. Matter} \textbf{\bibinfo{volume}{19}},
  \bibinfo{pages}{255207} (\bibinfo{year}{2007}).

\bibitem[{\citenamefont{Zhu et~al.}(2000)\citenamefont{Zhu, Zheng, and
  Prinz}}]{cmuJAP2000}
\bibinfo{author}{\bibfnamefont{J.~G.} \bibnamefont{Zhu}},
  \bibinfo{author}{\bibfnamefont{Y.}~\bibnamefont{Zheng}}, \bibnamefont{and}
  \bibinfo{author}{\bibfnamefont{G.~A.} \bibnamefont{Prinz}},
  \bibinfo{journal}{J. Appl. Phys.}  (\bibinfo{year}{2000}).

\bibitem[{\citenamefont{Akerman}(2005)}]{MRAM}
\bibinfo{author}{\bibfnamefont{J.}~\bibnamefont{Akerman}},
  \bibinfo{journal}{Science} \textbf{\bibinfo{volume}{308}},
  \bibinfo{pages}{508} (\bibinfo{year}{2005}).

\bibitem[{\citenamefont{Muratov and Osipov}(2008)}]{360domain2}
\bibinfo{author}{\bibfnamefont{C.~B.} \bibnamefont{Muratov}} \bibnamefont{and}
  \bibinfo{author}{\bibfnamefont{V.~V.} \bibnamefont{Osipov}},
  \emph{\bibinfo{title}{arxiv:0811.4663v1}} (\bibinfo{year}{2008}).

\bibitem[{\citenamefont{Cowburn and Welland}(2000)}]{Cowburnlogic}
\bibinfo{author}{\bibfnamefont{R.~P.} \bibnamefont{Cowburn}} \bibnamefont{and}
  \bibinfo{author}{\bibfnamefont{M.~E.} \bibnamefont{Welland}},
  \bibinfo{journal}{Science} \textbf{\bibinfo{volume}{287}},
  \bibinfo{pages}{1466} (\bibinfo{year}{2000}).

\bibitem[{\citenamefont{Prinz}(1998)}]{sensor1}
\bibinfo{author}{\bibfnamefont{G.~A.} \bibnamefont{Prinz}},
  \bibinfo{journal}{Science} \textbf{\bibinfo{volume}{282}},
  \bibinfo{pages}{1660} (\bibinfo{year}{1998}).

\bibitem[{\citenamefont{Miller et~al.}(2002)\citenamefont{Miller, Prinz, Cheng,
  and Bounnak}}]{sensor2}
\bibinfo{author}{\bibfnamefont{M.~M.} \bibnamefont{Miller}},
  \bibinfo{author}{\bibfnamefont{G.~A.} \bibnamefont{Prinz}},
  \bibinfo{author}{\bibfnamefont{S.~F.} \bibnamefont{Cheng}}, \bibnamefont{and}
  \bibinfo{author}{\bibfnamefont{S.}~\bibnamefont{Bounnak}},
  \bibinfo{journal}{Appl. Phys. Lett.} \textbf{\bibinfo{volume}{81}},
  \bibinfo{pages}{2211} (\bibinfo{year}{2002}).

\bibitem[{\citenamefont{Kl$\ddot{a}ui$}(2008)}]{KlauiReview2008}
\bibinfo{author}{\bibfnamefont{M.}~\bibnamefont{Kl$\ddot{a}ui$}},
  \bibinfo{journal}{J. Phys.: Condens. Matter} \textbf{\bibinfo{volume}{20}},
  \bibinfo{pages}{313001} (\bibinfo{year}{2008}).

\bibitem[{\citenamefont{Zhang et~al.}(2008)\citenamefont{Zhang, Singh,
  Bray-Ali, and Haas}}]{me}
\bibinfo{author}{\bibfnamefont{W.}~\bibnamefont{Zhang}},
  \bibinfo{author}{\bibfnamefont{R.}~\bibnamefont{Singh}},
  \bibinfo{author}{\bibfnamefont{N.}~\bibnamefont{Bray-Ali}}, \bibnamefont{and}
  \bibinfo{author}{\bibfnamefont{S.}~\bibnamefont{Haas}},
  \bibinfo{journal}{Phys. Rev. B} \textbf{\bibinfo{volume}{77}},
  \bibinfo{pages}{144428} (\bibinfo{year}{2008}).

\bibitem[{\citenamefont{Singh et~al.}(2008)\citenamefont{Singh, Krotkov, Xiang,
  Xu, Russell, and Tuominen}}]{ultrasmall1}
\bibinfo{author}{\bibfnamefont{D.~K.} \bibnamefont{Singh}},
  \bibinfo{author}{\bibfnamefont{R.~V.} \bibnamefont{Krotkov}},
  \bibinfo{author}{\bibfnamefont{H.}~\bibnamefont{Xiang}},
  \bibinfo{author}{\bibfnamefont{T.}~\bibnamefont{Xu}},
  \bibinfo{author}{\bibfnamefont{T.~P.} \bibnamefont{Russell}},
  \bibnamefont{and} \bibinfo{author}{\bibfnamefont{M.~T.}
  \bibnamefont{Tuominen}}, \bibinfo{journal}{Nanotechnology}
  \textbf{\bibinfo{volume}{19}}, \bibinfo{pages}{245305}
  (\bibinfo{year}{2008}).

\bibitem[{\citenamefont{Chaves-O'Flynn
  et~al.}(2008)\citenamefont{Chaves-O'Flynn, Kent, and Stein}}]{saddle}
\bibinfo{author}{\bibfnamefont{G.~D.} \bibnamefont{Chaves-O'Flynn}},
  \bibinfo{author}{\bibfnamefont{A.~D.} \bibnamefont{Kent}}, \bibnamefont{and}
  \bibinfo{author}{\bibfnamefont{D.~L.} \bibnamefont{Stein}},
  \emph{\bibinfo{title}{arxiv:0811.4440v1}} (\bibinfo{year}{2008}).

\bibitem[{\citenamefont{Beleggia et~al.}(2006)\citenamefont{Beleggia, Lau,
  Schofield, Zhu, Tandon, and DeGraef}}]{nanoring1}
\bibinfo{author}{\bibfnamefont{M.}~\bibnamefont{Beleggia}},
  \bibinfo{author}{\bibfnamefont{J.~W.} \bibnamefont{Lau}},
  \bibinfo{author}{\bibfnamefont{M.~A.} \bibnamefont{Schofield}},
  \bibinfo{author}{\bibfnamefont{Y.}~\bibnamefont{Zhu}},
  \bibinfo{author}{\bibfnamefont{S.}~\bibnamefont{Tandon}}, \bibnamefont{and}
  \bibinfo{author}{\bibfnamefont{M.}~\bibnamefont{DeGraef}},
  \bibinfo{journal}{J. Magn. Magn. Mater.} \textbf{\bibinfo{volume}{301}},
  \bibinfo{pages}{131} (\bibinfo{year}{2006}).

\bibitem[{\citenamefont{Landeros et~al.}(2006)\citenamefont{Landeros, Escrig,
  Altbir, Bahiana, and d'Albuquerque~e Castro}}]{nanoring2}
\bibinfo{author}{\bibfnamefont{P.}~\bibnamefont{Landeros}},
  \bibinfo{author}{\bibfnamefont{J.}~\bibnamefont{Escrig}},
  \bibinfo{author}{\bibfnamefont{D.}~\bibnamefont{Altbir}},
  \bibinfo{author}{\bibfnamefont{M.}~\bibnamefont{Bahiana}}, \bibnamefont{and}
  \bibinfo{author}{\bibfnamefont{J.}~\bibnamefont{d'Albuquerque~e Castro}},
  \bibinfo{journal}{J. Appl. Phys.} \textbf{\bibinfo{volume}{100}},
  \bibinfo{pages}{044311} (\bibinfo{year}{2006}).

\bibitem[{\citenamefont{Laufenberg et~al.}(2006)\citenamefont{Laufenberg,
  Backes, Buhrer, Bedau, Kl$\ddot{a}$ui, and et~al}}]{KlauiAPL2006}
\bibinfo{author}{\bibfnamefont{M.}~\bibnamefont{Laufenberg}},
  \bibinfo{author}{\bibfnamefont{D.}~\bibnamefont{Backes}},
  \bibinfo{author}{\bibfnamefont{W.}~\bibnamefont{Buhrer}},
  \bibinfo{author}{\bibfnamefont{D.}~\bibnamefont{Bedau}},
  \bibinfo{author}{\bibfnamefont{M.}~\bibnamefont{Kl$\ddot{a}$ui}},
  \bibnamefont{and} \bibinfo{author}{\bibnamefont{et~al}},
  \bibinfo{journal}{Appl. Phys. Lett.} \textbf{\bibinfo{volume}{88}},
  \bibinfo{pages}{052507} (\bibinfo{year}{2006}).

\bibitem[{\citenamefont{Zhang and Haas}(2008)}]{meReview}
\bibinfo{author}{\bibfnamefont{W.}~\bibnamefont{Zhang}} \bibnamefont{and}
  \bibinfo{author}{\bibfnamefont{S.}~\bibnamefont{Haas}},
  \emph{\bibinfo{title}{arxiv:0902.3024v1}} (\bibinfo{year}{2008}).

\bibitem[{\citenamefont{Kl$\ddot{a}ui$
  et~al.}(2005)\citenamefont{Kl$\ddot{a}ui$, Vaz, Heyderman, and
  Bland}}]{KlauiJMMM2005}
\bibinfo{author}{\bibfnamefont{M.}~\bibnamefont{Kl$\ddot{a}ui$}},
  \bibinfo{author}{\bibfnamefont{C.~A.~F.} \bibnamefont{Vaz}},
  \bibinfo{author}{\bibfnamefont{L.~J.} \bibnamefont{Heyderman}},
  \bibnamefont{and} \bibinfo{author}{\bibfnamefont{J.~A.~C.}
  \bibnamefont{Bland}}, \bibinfo{journal}{J. Magn. Magn. Mater.}
  \textbf{\bibinfo{volume}{290}}, \bibinfo{pages}{61} (\bibinfo{year}{2005}).

\bibitem[{\citenamefont{Castano et~al.}(2003)\citenamefont{Castano, Ross,
  Frandsen, Eilez, gil, Smith, Redjdal, and Humphrey}}]{RossPRB2003}
\bibinfo{author}{\bibfnamefont{F.~J.} \bibnamefont{Castano}},
  \bibinfo{author}{\bibfnamefont{C.~A.} \bibnamefont{Ross}},
  \bibinfo{author}{\bibfnamefont{C.}~\bibnamefont{Frandsen}},
  \bibinfo{author}{\bibfnamefont{A.}~\bibnamefont{Eilez}},
  \bibinfo{author}{\bibfnamefont{D.}~\bibnamefont{gil}},
  \bibinfo{author}{\bibfnamefont{H.~I.~.} \bibnamefont{Smith}},
  \bibinfo{author}{\bibfnamefont{M.}~\bibnamefont{Redjdal}}, \bibnamefont{and}
  \bibinfo{author}{\bibfnamefont{F.~B.} \bibnamefont{Humphrey}},
  \bibinfo{journal}{Phys. Rev. B} \textbf{\bibinfo{volume}{67}},
  \bibinfo{pages}{184425} (\bibinfo{year}{2003}).

\bibitem[{\citenamefont{Moore et~al.}(2005)\citenamefont{Moore, Hayward, Tse,
  Bland, Castano, and Ross}}]{RossJAP2005}
\bibinfo{author}{\bibfnamefont{T.~A.} \bibnamefont{Moore}},
  \bibinfo{author}{\bibfnamefont{T.~J.} \bibnamefont{Hayward}},
  \bibinfo{author}{\bibfnamefont{D.~H.~Y.} \bibnamefont{Tse}},
  \bibinfo{author}{\bibfnamefont{J.~A.~C.} \bibnamefont{Bland}},
  \bibinfo{author}{\bibfnamefont{F.~J.} \bibnamefont{Castano}},
  \bibnamefont{and} \bibinfo{author}{\bibfnamefont{C.~A.} \bibnamefont{Ross}},
  \bibinfo{journal}{J. Appl. Phys.} \textbf{\bibinfo{volume}{97}},
  \bibinfo{pages}{063910} (\bibinfo{year}{2005}).

\bibitem[{\citenamefont{Singh et~al.}(2009)\citenamefont{Singh, Krotkov, and
  Tuominen}}]{ultrasmall2}
\bibinfo{author}{\bibfnamefont{D.~K.} \bibnamefont{Singh}},
  \bibinfo{author}{\bibfnamefont{R.~V.} \bibnamefont{Krotkov}},
  \bibnamefont{and} \bibinfo{author}{\bibfnamefont{M.~T.}
  \bibnamefont{Tuominen}}, \bibinfo{journal}{Phys. Rev. B}
  \textbf{\bibinfo{volume}{79}}, \bibinfo{pages}{184409}
  (\bibinfo{year}{2009}).

\bibitem[{\citenamefont{d'Albuquerque~e Castro
  et~al.}(2002)\citenamefont{d'Albuquerque~e Castro, Altbir, Retamal, and
  Vargas}}]{scale}
\bibinfo{author}{\bibfnamefont{J.}~\bibnamefont{d'Albuquerque~e Castro}},
  \bibinfo{author}{\bibfnamefont{D.}~\bibnamefont{Altbir}},
  \bibinfo{author}{\bibfnamefont{J.~C.} \bibnamefont{Retamal}},
  \bibnamefont{and} \bibinfo{author}{\bibfnamefont{P.}~\bibnamefont{Vargas}},
  \bibinfo{journal}{Phys. Rev. Lett.} \textbf{\bibinfo{volume}{88}},
  \bibinfo{pages}{237202} (\bibinfo{year}{2002}).

\bibitem[{\citenamefont{Mejia-Lopez et~al.}(2006)\citenamefont{Mejia-Lopez,
  Altbir, Romero, Batlle, Roshchin, Li, and Schuller}}]{mc1}
\bibinfo{author}{\bibfnamefont{J.}~\bibnamefont{Mejia-Lopez}},
  \bibinfo{author}{\bibfnamefont{D.}~\bibnamefont{Altbir}},
  \bibinfo{author}{\bibfnamefont{A.~H.} \bibnamefont{Romero}},
  \bibinfo{author}{\bibfnamefont{X.}~\bibnamefont{Batlle}},
  \bibinfo{author}{\bibfnamefont{I.~V.} \bibnamefont{Roshchin}},
  \bibinfo{author}{\bibfnamefont{C.}~\bibnamefont{Li}}, \bibnamefont{and}
  \bibinfo{author}{\bibfnamefont{I.~K.} \bibnamefont{Schuller}},
  \bibinfo{journal}{J. Appl. Phys.} \textbf{\bibinfo{volume}{100}},
  \bibinfo{pages}{104319} (\bibinfo{year}{2006}).

\bibitem[{\citenamefont{Mejia-Lopez et~al.}(2005)\citenamefont{Mejia-Lopez,
  Soto, and Altbir}}]{mc2}
\bibinfo{author}{\bibfnamefont{J.}~\bibnamefont{Mejia-Lopez}},
  \bibinfo{author}{\bibfnamefont{P.}~\bibnamefont{Soto}}, \bibnamefont{and}
  \bibinfo{author}{\bibfnamefont{D.}~\bibnamefont{Altbir}},
  \bibinfo{journal}{Phys. Rev. B} \textbf{\bibinfo{volume}{71}},
  \bibinfo{pages}{104422} (\bibinfo{year}{2005}).

\bibitem[{\citenamefont{Zhang and Haas}(2009)}]{mefmm}
\bibinfo{author}{\bibfnamefont{W.}~\bibnamefont{Zhang}} \bibnamefont{and}
  \bibinfo{author}{\bibfnamefont{S.}~\bibnamefont{Haas}}, \bibinfo{journal}{J.
  Magn. Magn. Mater.} \textbf{\bibinfo{volume}{321}}, \bibinfo{pages}{3687}
  (\bibinfo{year}{2009}).

\bibitem[{\citenamefont{Nowak et~al.}(2000)\citenamefont{Nowak, Chantrell, and
  Kennedy}}]{mc3}
\bibinfo{author}{\bibfnamefont{N.}~\bibnamefont{Nowak}},
  \bibinfo{author}{\bibfnamefont{R.~W.} \bibnamefont{Chantrell}},
  \bibnamefont{and} \bibinfo{author}{\bibfnamefont{E.~C.}
  \bibnamefont{Kennedy}}, \bibinfo{journal}{Phys. Rev. Lett.}
  \textbf{\bibinfo{volume}{84}}, \bibinfo{pages}{163} (\bibinfo{year}{2000}).

\bibitem[{\citenamefont{Li et~al.}(2001)\citenamefont{Li, Peyrade, Natali,
  Lebib, Chen, Ebels, Buda, and Ounadjela}}]{francePRL2001}
\bibinfo{author}{\bibfnamefont{S.~P.} \bibnamefont{Li}},
  \bibinfo{author}{\bibfnamefont{D.}~\bibnamefont{Peyrade}},
  \bibinfo{author}{\bibfnamefont{M.}~\bibnamefont{Natali}},
  \bibinfo{author}{\bibfnamefont{A.}~\bibnamefont{Lebib}},
  \bibinfo{author}{\bibfnamefont{Y.}~\bibnamefont{Chen}},
  \bibinfo{author}{\bibfnamefont{U.}~\bibnamefont{Ebels}},
  \bibinfo{author}{\bibfnamefont{L.~D.} \bibnamefont{Buda}}, \bibnamefont{and}
  \bibinfo{author}{\bibfnamefont{K.}~\bibnamefont{Ounadjela}},
  \bibinfo{journal}{Phys. Rev. Lett.} \textbf{\bibinfo{volume}{86}},
  \bibinfo{pages}{1102} (\bibinfo{year}{2001}).

\bibitem[{\citenamefont{Steiner and Nitta}(2004)}]{triple}
\bibinfo{author}{\bibfnamefont{M.}~\bibnamefont{Steiner}} \bibnamefont{and}
  \bibinfo{author}{\bibfnamefont{J.}~\bibnamefont{Nitta}},
  \bibinfo{journal}{Appl. Phys. Lett.} \textbf{\bibinfo{volume}{84}},
  \bibinfo{pages}{939} (\bibinfo{year}{2004}).

\bibitem[{\citenamefont{Kl$\ddot{a}$ui and et~al}(2004)}]{KlauiAPL2004}
\bibinfo{author}{\bibfnamefont{M.}~\bibnamefont{Kl$\ddot{a}$ui}}
  \bibnamefont{and} \bibinfo{author}{\bibnamefont{et~al}},
  \bibinfo{journal}{Appl. Phys. Lett.} \textbf{\bibinfo{volume}{84}},
  \bibinfo{pages}{951} (\bibinfo{year}{2004}).

\bibitem[{\citenamefont{Steiner et~al.}(2004)\citenamefont{Steiner, Meier,
  Merkt, and Nitta}}]{triple2}
\bibinfo{author}{\bibfnamefont{M.}~\bibnamefont{Steiner}},
  \bibinfo{author}{\bibfnamefont{G.}~\bibnamefont{Meier}},
  \bibinfo{author}{\bibfnamefont{U.}~\bibnamefont{Merkt}}, \bibnamefont{and}
  \bibinfo{author}{\bibfnamefont{J.}~\bibnamefont{Nitta}},
  \bibinfo{journal}{Physica E}  (\bibinfo{year}{2004}).

\bibitem[{\citenamefont{Yan et~al.}(2007)\citenamefont{Yan, Hertel, and
  Schneider}}]{vertical}
\bibinfo{author}{\bibfnamefont{M.}~\bibnamefont{Yan}},
  \bibinfo{author}{\bibfnamefont{R.}~\bibnamefont{Hertel}}, \bibnamefont{and}
  \bibinfo{author}{\bibfnamefont{C.~M.} \bibnamefont{Schneider}},
  \bibinfo{journal}{Phys. Rev. B} \textbf{\bibinfo{volume}{76}},
  \bibinfo{pages}{094407} (\bibinfo{year}{2007}).

\end{thebibliography}

\end{document}